\documentstyle[preprint,aps,eqsecnum]{revtex}
 
\input{psfig}
\def\be{\begin{equation}}
\def\ee{\end{equation}}
\def\bea{\begin{eqnarray}} 
\def\eea{\end{eqnarray}}
\def\line{\hbox to \hsize}    
\def\frac #1#2{{#1\over #2}}

\def\hH{{\hat H}}

\def \ket #1{{\vert #1\rangle}}
\def \bra #1{{\langle #1\vert}}

\def\eval #1#2#3{{\langle#1\vert#2\vert#3\rangle}} 

\def\1{\mbox{\bf 1}} 

\begin{document}
\draft 

\title{
A note on the time evolution of generalized  coherent states
}

\author{MICHAEL STONE}
\address{University of Illinois, Department of Physics\\ 1110 W. Green St.\\
Urbana, IL 61801 USA\\E-mail: m-stone5@uiuc.edu} 
\maketitle

\begin{abstract}

I consider the  time evolution of generalized coherent states based on
non-standard fiducial vectors, and  show
that only for a restricted class of fiducial vectors does
the associated  classical motion determine the quantum
evolution of the states. I discuss some  consequences of this for path
integral representations.

\end{abstract}


\section{Introduction}

Coherent states were originally introduced into physics by
Schr{\"o}dinger\cite{schrodinger26} in order  to reconcile
Heisenberg's  abstract solution of the quantum  harmonic
oscillator with the  classical picture of a swinging
pendulum. Schr{\"o}dinger's minimum-uncertainty 
wave-packets maintain their shape under the quantum time
evolution and   their ``centre-of-mass'' motion coincides
with that of the pendulum.  There are various
generalizations of the original harmonic oscillator
coherent states, but most of them exhibit the following 
classical-quantum correspondence:

\begin{itemize}

\item There is a 
family of states parameterized by some classically
interpretable variables.

\item For  simple Hamiltonians, a state
originally in  this family remains in it.   

\item  The parameters evolve according to the classical equations of
motion. 

\item In  addition to the
parameter evolution, the states  accumulate  a phase equal
to the exponential of the classical action.

\end{itemize}

The last property was  pointed out by 
van Hove\cite{vanHove51}.  If all four of these
features are present, the classical dynamics completely
determines the quantum time evolution.       

One such extension of the Schr{\"o}dinger states is the
class of  ``generalized coherent states'' introduced by  
Perelomov\cite{perelomov72,perelomov86}. Perelomov's  states are
obtained by selecting a  fiducial vector in some 
representation of a Lie group, and considering its orbit
under the action of the group. For compact groups (and for
some representations of non-compact groups) this
construction guarantees the existence of a resolution of
unity, and so suggests a path integral representation of
the dynamics\cite{klauder79,kuratsuji80}.

Usually the Perelomov  coherent states are constructed by
taking a greatest (or least) weight state as the fiducial
vector. This generates  minimum
uncertainty  states and, for Hamiltonians that are elements
of the Lie algebra, the states evolve  classically in the
manner described above\cite{onofri75}.   In  recent
articles, however,
M.~Matsumoto\cite{matsumoto96,matsumoto99} has considered 
coherent states built on arbitrary
elements of the  representation space.   His motivation is 
that such states are  useful in quantum optics
and that novel  phase interference effects  may be
measurable.

The purpose of this note is to  point out what 
must be well known---although I have not been able to find an
explicit statement in the literature---that these
``general'' generalized coherent states {\it do not\/}
necessarily  inherit the classical-quantum correspondence 
property.  If we insist on the traditional picture, then we
must restrict our choice of fiducial vectors. The essential
requirement is   that  two distinct notions  of the 
isotropy group associated with the fiducial state must,
in the end, define the same group.

In the next section I first review some basic facts about
Perelomov coherent states and  define  the two isotropy
groups associated with a fiducial vector.  In the third
section I show that only if these two isotropy groups
coincide does the classical motion provide complete
information about the quantum dynamics.     Finally I make
some brief remarks on the implication of these issues for 
path integrals.

\section{Informative families}

Suppose $G$ is a compact Lie group and $\ket{0}$ a vector
in an  irreducible representation space, $V$, of $G$.
Since $G$ is compact,  we may assume that $V$ comes   equipped with an
inner product with respect to which the representation is 
unitary.
For  $g\in G$  an element of the group we will write its
action on a vector $\ket{v}\in V $ by $\ket{v}\to
g\ket{v}$. Thus we make no notational distinction between
the elements of the group and the corresponding  operators
$g:V\to V$ in the representation.   We will also write
$\ket{g}=g\ket{0}$.  Following Perelomov\cite{perelomov86},
we will call the  set  $\{\ket{g}\}$  a family of {\it
generalized coherent states\/}. The starting  $\ket{0}$ is 
the {\it fiducial vector \/}  of the family.

If $d\mu$ denotes the  invariant  measure on the group then
\be
B=\int d\mu \ket{g}\bra{g}
\ee commutes with all matrices $g$, and so, by Shur's lemma,
is proportional to the identity operator. Thus any $\ket{0}$
provides a resolution of the identity
\be
{\bf I}= const. \int d\mu \ket{g}\bra{g}.   
\ee
This is probably the  most important property of such sets
of coherent states.

Although any state in the representation produces a family 
of states with resolution of the identity,  the  families
are not all  equivalent, nor are all  equally
useful.  The most commonly seen  are those built on highest
(or lowest) weight vectors, such as the state $\ket{j,j}$
in the spin $j$ representation of $SU(2)$. As with the
Schr{\"o}dinger wave-packets, these  families  are composed
of  
minimum uncertainty  states, and have other nice
mathematical properties. In particular they  are naturally 
complex homogeneous manifolds with a  K{\"a}hler
structure\cite{onofri75}. Here, however, we  are interested
in  a broader class of fiducial vectors.

For any $\ket{0}$ consider the following sets:

\begin{itemize}

\item $H_{\ket{0}} =\{  h\in G|\,  h \ket{0}
= ({\rm phase})\ket{0}\}$ --- {\it i.e.\/}   the set of elements of
$G$ for which $\ket{0}$ is a common eigenvector.

\item $H_0 =\{  h\in G|\, \eval{ h}{\hat \lambda}{
h}= \eval{0}{\hat \lambda}{0}\,
 \forall \hat \lambda \in {\rm Lie\,}(G)\}$ --- {\it i.e.\/} the set of elements of
$G$ which stabilizes $f_{ 0} =\eval{0}{\ldots }{0}$ considered as an element of
$\left({\rm Lie\,}(G)\right)^*$.

\end{itemize}

Clearly both sets are subgroups and  $ H_{\ket{0}}\subseteq H_0$. 

The first subgroup, $H_{\ket{0}}$, is the isotropy group of the
family of coherent states. In other words, the physically
distinct states in the family are in one-to-one correspondence with the
quotient space $G/H_{\ket{0}}$. States in any particular
coset differ only by an overall phase.

The second group, $H_0$, is the isotropy group of the
linear functional  \hbox{$f_0: {\rm Lie\,}(G)
\to {\bf C}$}, where $f_0(\hat\lambda) =\eval{0}{\hat \lambda}{0}$, 
under the co-adjoint action of $G$ {\it i.e.\/}
\be
f_0(\hat\lambda) \to f_g(\hat\lambda) = f_0(g^{-1}\hat \lambda g).
\ee
As we will see, the degrees of freedom of the 
associated classical system
live in  the co-adjoint orbit  $G/H_0$.  

Since we wish the classical and quantum evolutions to
be related, it would be nice if the two coset spaces
$G/H_{\ket{0}}$ and  $G/H_0$ were the same.
Unfortunately the  two subgroups  $H_{\ket{0}}$ and $H_0$ do
not necessarily  
coincide.

As an example, consider one of the families  discussed by
Matsumoto\cite{matsumoto99}. We take as fiducial vector the state
 \be
\ket{0}= \sqrt{{\frac 23}} \ket{1,1} + \sqrt{{\frac 13}}
\ket{1,-1}
\label{EQ:matsumotoket}
\ee  
in the spin-$1$ representation of $SU(2)$.
We have 
\bea
\eval{0}{\hat J_1}{0}&=& 0\nonumber\\
\eval{0}{\hat J_2}{0}&=& 0\nonumber\\ 
\eval{0}{\hat J_3}{0}&=& \frac 13.
\eea
The isotropy group $H_0$ consists of rotations
about the ``$3$'' axis. The group $H_{\ket{0}}$
must be a subgroup of this, but under such rotations
\be
\sqrt{{\frac 23}} \ket{1,1} + \sqrt{{\frac 13}}
\ket{1,-1}\to  e^{i\theta}\sqrt{{\frac 23}} \ket{1,1} +
e^{-i\theta}\sqrt{{\frac 13}}
\ket{1,-1}.
\label{EQ:phaseshift}
\ee
Thus $H_0=\{e^{i\theta J_3}\}$,  while $H_{\ket{0}}$ contains
only the identity element.

Suppose now that $H_0$ {\it is\/} equal to $H_{\ket{0}}$. 
In this case the cosets 
$G/H_{\ket{0}}$ and  $G/H_0$ coincide and consequently  $\ket{g}$ is
determined (up to a phase)  by the values of the
expectations $\eval{g}{\hat \lambda_i}{g} =\lambda_i$ as
$\hat \lambda_i$ ranges over the Lie algebra of $G$.  This
latter property is a reasonable one to require, since then
we can determine (up to a phase) the quantum state of the system from
measurement of the these expectations.   There may be a
standard name for fiducial vectors for which
$H_0=H_{\ket{0}}$, but I am unaware of it. For want of a
suitable word, I will call such vectors, and 
the resulting family of coherent states,
{\it informative\/}.

\section{Dynamics}

If we restrict our attention to Hamiltonians that are
elements of the Lie algebra, then a state that starts as a
coherent state will remain one under the quantum time
evolution.  

To associate a classical dynamical process with the quantum
one, let $\hH\in {\rm Lie}(G)$ be the quantum Hamiltonian,
$\ket{0}$ be our selected fiducial vector, 
and define the action 
\be
S= \int\, dt \left\{ i\eval{0}{g^{-1}\dot
g}{0} - \eval{0}{g^{-1}\hH g}{0}\right\}.
\label{EQ:classaction}
\ee  
The first term is the  canonical (or Berry) phase which
determines the symplectic structure on the phase
space\cite{kirillov76,kostant70a,kostant70b}. The second term serves as the
classical Hamiltonian.

The variation of $S$ is  
\be 
\delta S= 
\int\, dt \left\{ \eval{0}{[ig^{-1}\dot g -
g^{-1}\hH g, g^{-1}\delta g]}{0}\right\}.
\ee
The corresponding equation of motion for $g$ is therefore
\be
g^{-1}\dot g=  -i
g^{-1}\hH g +i\hat \lambda(t),
\label{EQ:eqofmot}
\ee
where $\hat \lambda(t)$ is any element of the Lie algebra
obeying 
$\eval{0}{[i\hat \lambda(t), g^{-1} \delta g]}{0}=0$ for all 
$g^{-1} \delta g\in {\rm Lie}(G)$. This  condition
means that $\hat \lambda(t)$ lies in  the Lie algebra of $H_0$. 
Since  $g^{-1}\dot g$ is indeterminate up to an an
element of ${\rm Lie\,}(H_0)$, we must consider the
classical trajectories as living in the co-adjoint orbit $G/H_0$,
rather than in $G$. In general, knowing the classical
trajectory in this space is not enough to determine the
evolution of the quantum state. For example, the evolution of
the fiducial state (\ref{EQ:matsumotoket}) with $\hat H=\hat
J_3$ alters  the relative phase of the two components as in 
(\ref{EQ:phaseshift}). This phase-shift is invisible in the classical
motion since the expectation values of the Lie algebra
generators do not change.

We can make further progress, however,  if  we take  
$\ket{0}$ to be  informative. This means $\hat \lambda(t)$ is
also in $H_{\ket{0}}$, so 
\be
\hat\lambda(t)\ket{0}=\lambda(t) \ket{0}
\ee
for some (real) number $\lambda(t)$. 

Given this information we   note that  
\be
S= -\int\lambda(t)dt.
\ee
Next we observe that the solution of the equation of motion
for $g(t)$ is
\be
g(t) =T\left\{e^{-i  \int_0^t \hH dt}\right\}
g(0) \overline{T}\left\{e^{i\int_0^t \hat\lambda
dt}\right\},
\ee
where $\overline{T}$ denotes anti-time-ordering.
Therefore
\be
\ket{g(t)}=g(t)\ket{0}= T\left\{e^{-i
\int_0^t \hH dt}\right\}g(0) e^{i\int_0^t\lambda(t) dt}\ket{0}.
\ee  

In other words
\be
T\left\{e^{-i
\int_0^t \hH dt}\right\}\ket{g(0)}=e^{ iS}\ket{g(t)}.
\ee
The left hand side of this equation is the quantum time-evolved 
coherent state, while the right hand side is, up
to a phase, the coherent state corresponding to the
classically  evolved  variable $g(t)$.     

At first sight the action $S$ appearing in the above
expression is arbitrary. This is because in equation 
(\ref{EQ:eqofmot})  $\hat \lambda$ could have been any
element of ${\rm Lie\,}(H_0)$. The ambiguity is removed
however, when we select a   specific  representative,
$\ket{g}$, from each ray in $G/H_{\ket{0}}$. This is what
is normally done when we define a family of coherent
states. For informative states, we have therefore recovered
the traditional picture dating back to
van~Hove\cite{vanHove51}. This result, that the quantum
evolution may be found by solving a purely classical
problem, is essentially  equivalent to the Wei-Norman disentangling
procedure\cite{weinorman63}.

\section{Discussion}

Because {\it any} fiducial vector gives rise to a
resolution of unity, we can use the coherent states
constructed on it to write down an exact discrete-time path
integral for any transition amplitude.  From this, by
taking a formal limit of infinitely many intermediate
steps,  we may  ``derive'' a  continuous-time path integral
representation of the quantum dynamics. The action appearing
in this path integral is (\ref{EQ:classaction}), together
with some boundary terms that serve to make the initial and
final value problem well defined. We might
reasonably  expect the classical paths, those with
stationary variations, to play an important role in
evaluating this path integral. Unfortunately we have seen
that  for the general fiducial vector  these paths do not
capture the full quantum dynamics. We must expect,
therefore, substantial  analytic difficulties in making
rigorous the  continuous-time  limit of the discrete path
integral.       

As a symptom of these problems, consider the formal path
integral that comes from taking (\ref{EQ:matsumotoket}) as
fiducial vector. The   canonical phase term in the
classical action is then \cite{matsumoto99}   
\be
\int \frac {1}{3} (\cos \theta - 1)\dot \phi\, dt.    
\ee
The coefficient, $1/3$, violates the  condition
required to  make the ``Dirac string'' at the south pole
invisible. Only integers and half integers are allowed as
coefficients if the path integral is to be
well-defined\cite{witten83a}. 

For the spin coherent states of $SU(2)$, it has recently been
shown\cite{stone00} that, when  the fiducial vector is
taken to be a highest weight vector,  the formal  semi-classical
expansion about the continuous-time classical paths does
yield  correct answers. It would be a salutary exercise  to trace 
exactly what goes wrong with the continuous-time limit in
the more general case.

\acknowledgements

This work was supported by the National Science Foundation
under grant DMR-98-17941. I  would also like to thank  the
TCM group at the Cavendish Laboratory, Cambridge, England,
for hospitality, and the  EPSRC for funding my visit under
grant number GR/N00364.

\eject
\end{document}